\documentclass[runningheads]{llncs}

\usepackage{url}
\usepackage{bbold}
\usepackage{multirow}
\usepackage{graphicx, nicefrac}
\usepackage{booktabs}
\usepackage{amsfonts}
\usepackage{amsmath}
\usepackage{enumitem}
\usepackage{comment}

\usepackage[sort]{cite}
\usepackage[hidelinks]{hyperref}
\usepackage[misc]{ifsym}
\usepackage{xcolor}
\usepackage[linesnumbered,vlined,ruled]{algorithm2e}
\usepackage{orcidlink}

\usepackage{tikz}
\usetikzlibrary{decorations.pathreplacing,calc}
\newcommand{\tikzmark}[1]{\tikz[overlay,remember picture] \node (#1) {};}

\newcommand*{\AddNoteA}[5]{%
    \begin{tikzpicture}[overlay, remember picture]
        \draw [ultra thick, #4]
            ($(#3)!(#1.north)!($(#3)-(0,1)$)$) --  ($(#3)!(#2.south)!($(#3)-(0,1)$)$)
                node [align=center, text width=2cm, pos=0.5, anchor=west, rotate=90, anchor=north, yshift=5mm, fill=black!1, opacity=0.8] {#5};
    \end{tikzpicture}
}%

\newcommand{\frank}{\textsc{Frank}}

\definecolor{darkgreen}{rgb}{0.0, 0.5, 0.0}
\definecolor{darkblue}{rgb}{0.0, 0.2, 0.6}
\definecolor{darkpastelred}{rgb}{0.76, 0.23, 0.13}
\definecolor{carrotorange}{rgb}{0.93, 0.57, 0.13}

\begin{document}

\setlength{\abovedisplayskip}{3pt}
\setlength{\belowdisplayskip}{3pt}

\title{A Frank System for\\ Co-Evolutionary Hybrid Decision-Making}
\titlerunning{A Frank System for Co-Evolutionary Hybrid Decision-Making}

\author{Federico Mazzoni\inst{1}{\footnotesize{\Letter}} \orcidlink{0009-0004-6961-7495} \and
    Riccardo Guidotti\inst{2} \orcidlink{0000-0002-2827-7613} \and
    Alessio Malizia\inst{3} \orcidlink{0000-0002-2601-7009} 
}

\authorrunning{F. Mazzoni et al.}


\institute{
    $^1 {^2 {^3}}$University of Pisa, Italy, $^3$Molde University College, Norway\\ 
    $^1$\email{\{name.surname\}@phd.unipi.it},
    $^2 {^3}$\email{\{name.surname\}@unipi.it}
}


\authorrunning{F. Mazzoni et al.}

\maketitle

\begin{abstract}
We introduce \frank{}, a human-in-the-loop system for co-evolutionary hybrid decision-making aiding the user to label records from an un-labeled dataset. 
\frank{} employs incremental learning to ``evolve'' in parallel with the user's decisions, by training an interpretable machine learning model on the records labeled by the user. 
Furthermore, \frank{} advances state-of-the-art approaches by offering inconsistency controls, explanations, fairness checks, and bad-faith safeguards simultaneously.
We evaluate our proposal by simulating the users' behavior with various levels of expertise and reliance on \frank{}'s suggestions. 
The experiments show that \frank{}'s intervention leads to improvements in the accuracy and the fairness of the decisions.

\end{abstract}

\section{Introduction}
Automated decision-makers based on Machine Learning (ML) are still not widely adopted for high-stakes decisions such as medical diagnoses or court decisions~\cite{wang2020augmented}.
In these fields, humans are aided but not replaced by Artificial Intelligence (AI), resulting in Hybrid Decision-Makers (HDM)~\cite{mosier2018human}.
While HDM literature is flourishing, certain key aspects have not yet been considered, preventing HDM systems from covering possible use cases.
HDM systems promote the collaboration between human and AI decision-makers, resulting in a final set of ``hybrid'' decisions (some taken by the human, others by the machine). 
In \textit{Learning-to-Defer}~\cite{joshi2021learning} systems, the machine plays the primary role, deferring decisions on records with a high degree of uncertainty to an external human supervisor. 
In~\cite{wang2020augmented}, a rule-based AI model with inferred rules suggests replacing some user's decisions to maximize fairness, whereas in~\cite{jarrett2022online}, the model mediates between a user and their supervisor if it is not confident in the user's decisions.
On the other hand, in the Skeptical Learning (SL) paradigm, an ML model learns ``in parallel'' to the decisions taken by a human and queries them if it is ``skeptical'' of the human decision~\cite{zhang2019personal,bontempelli2020learning,teso2021interactive,zeni2019fixing}.
SL aims to help the user remain consistent with their past decisions, still giving them veto power against the model's suggestions.
SL has been extensively applied to personal context recognition~\cite{bontempelli2020learning, zhang2019personal} and image classifications~\cite{teso2021interactive}. In~\cite{teso2021interactive}, \textsc{SL} suggestions are also supported by \textit{contrastive explanations}. 

Our system employs and extends traditional SL, by taking into account simultaneously fairness aspects, explainable suggestions, and the involvement of the user's supervisor. In line with~\cite{bontempelli2020learning}, our proposal is powered by a \textit{Incremental Learning} (IL) model. IL, also known as Continual Learning, is an ML paradigm where the model is continuously trained on small data batches, potentially including only one data point, instead of the entirety of the training set~\cite{de2021continual, wang2023comprehensive}.

The \textit{eXplainable AI (XAI)} research field aims to create humanly interpretable proxies of ``black-box'' ML models used for decision-making. 
An explanation is \textit{global} if it unveils the whole model logic, or \textit{local} if it justifies the decision of a specific record~\cite{guidotti2018survey}. 
A global explanation can be achieved by approximating black-box models with interpretable-by-design ones, such as a decision tree, which also offers local explanations as decision rules~\cite{blanco2019machine}. 
Also, instance-based explanations make use of examples and counter-examples, i.e., similar records with the same/different decision by the AI system~\cite{guidotti2022counterfactual}.
Our proposal offers both a model approximation, employing an interpretable decision tree, and instance-based (counter-)examples to explain the model's suggestions to the user.

Finally, we also account for 
the \textit{fairness} of the decisions.
Two major approaches have been proposed to quantify a dataset's fairness~\cite{binns2020apparent}. 
For \textit{individual fairness}, similar individuals should receive similar treatment, while for \textit{group fairness}, each group should receive a similar treatment~\cite{pessach2022review}.
The discriminatory feature to be monitored (e.g., \textit{Race}, \textit{Gender}) is often defined \textit{sensitive} or \textit{protected attribute}~\cite{verma2018fairness}.
Given a sensitive attribute, our proposal checks both individual and group fairness, helping the user avoid discriminating behavior.

We propose \frank{}\footnote{Original version of the paper available at: \url{https://link.springer.com/chapter/10.1007/978-3-031-58553-1_19}}, a HDM system overcoming the current limitations of \textsc{SL} related to explainability, fairness, consistency, and bad-faith users.
As in SL, if the user's label is inconsistent with \frank{}'s prediction, the user is warned of possible contradictions with their past behavior and suggested to modify their decision. 
Besides, \frank{} provides explanations that become increasingly detailed as the model learns more from the user, who can, in turn, learn more about their behavior. 
Also, \frank{} can prevent bad-faith behavior and discriminating decisions. 
Ultimately, \frank{} and the human have a symbiotic co-evolutionary relationship, with \frank{}'s model able to predict the user's behavior, thus aiding them, and the human feeding \frank{}'s model with new data.
Experimental results show that pairing \frank{} with less reliable users provides noticeable improvements in terms of accuracy and fairness, and that the usage of explanations increases the number of acceptance for suggestions in case of skepticism.

\section{Setting the Stage}
\label{sec:background}
We keep the paper self-contained by reporting in the following a brief overview of concepts necessary to understand our proposal. 
We indicate with $X, Y$ a dataset where $X = \{x_1, \dots, x_n\} \in \mathcal{X}^{(m)}$ is a set of $n$ records described by $m$ attributes (features), i.e., $x_i = \{(a_1, v_1), \dots, (a_m, v_m)\}$, where $a_i$ is the attribute name and $v_i$ is the corresponding value, and $\mathcal{X}^{(m)}$ is the feature space consisting of $m$ input features, while $Y = \{y_1, \dots, y_n\} \in \mathcal{Y}$ is the set of the target variable in the target space $\mathcal{Y}$.
With $A = \{a_1, \dots, a_m\}$ we indicate the set of feature names, and for an instance $x \in X$, we write $x[a_k]$ to refer to the value $v_k$ of attribute $a_k$.
For classification problems, $y_i \in \{1, \dots, l\} = L$ where $L$ is the set of different class labels and $l$ is the number of the classes, while when dealing with regression problems, $y_i \in \mathbb{R}$. 
Without losing in generality, we consider $l=2$, i.e., binary classification problems.
We indicate a trained decision-making model with a function $f:\mathcal{X}^{(m)} \rightarrow \mathcal{Y}$ that maps data instances $x$ from the feature space $\mathcal{X}^{(m)}$ to the target space $\mathcal{Y}$.
We write $f(x) = y$ to denote the decision $y$ taken by $f$, and $f(X) = Y$ as a shorthand for $\{f(x_i) \ \vert \ x_i \in X\} = Y$.

\smallskip
\textbf{Skeptical Learning.} 
Given a ML model $f$ and a dataset $X$, the user is tasked to assign a label $y_i$ to each record $x_i \in X$. 
In SL, the user assigns to $x_i$ the label $\hat{y}_i$, according to their own belief and background and, independently from them, $f$ assigns the label $\tilde{y}_i$, i.e, $\tilde{y}_i = f(x_i)$. 
The ML model implementing $f$ can be pre-trained on a small training set. 
If $\hat{y}_i \neq \tilde{y}_i$ and $f$ is \textit{skeptical} (see below), the user is asked if they want to accept $\tilde{y}_i$ as $y_i$. 
If they do, $y_i$ takes the value $\tilde{y}_i$. 
If the user refuses, if $\hat{y}_i = \tilde{y}_i$ or if the model is not skeptical, $y_i$ is assigned $\hat{y}_i$.
The ML model is then incrementally trained on $x_i$ and $y_i$.

The definition of the model's \textit{skepticality} varies in the literature~\cite{teso2021interactive}. 
However, skepticism is always related to model's \textit{epistemic uncertainty}, which is independent of the notion of \textit{confidence} score towards a certain decision, i.e., the prediction probability\footnote{\scriptsize Note that there's a general lack of normativity w.r.t. these terms; e.g.,~\cite{zeni2019fixing} uses the term \textit{confidence} to refer to the epistemic uncertainty.}. 
Epistemic uncertainty is the model's \textit{ignorance}, and given enough data, it should be minimized~\cite{hullermeier2021aleatoric}. Only a limited number of ML model offers by-design access to epistemic uncertainty, e.g., Naive Bayes, Gaussian Process~\cite{hullermeier2021aleatoric,bontempelli2020learning}. In the context of SL, it has been approximated by the \textit{empirical accuracy} of past predictions both of the user and the model, i.e., the ratio between the number of times a label has been proposed by the user or predicted by the model, and the times it has been accepted as $y$~\cite{zeni2019fixing}.
Thus, given $x_i$ and the prediction $\tilde{y}_i$, the skepticism towards the user's $\hat{y}_i$ is:
\begin{equation}
    \label{eq:skept}
    \mathit{skpt}(x_i, \tilde{y}_i, \hat{y}_i, Y, \tilde{Y}, \hat{Y}) = \mathit{c}(f, x_i, \tilde{y}_i) \cdot \mathit{ea}(\tilde{y}_i, Y, \tilde{Y}) - \mathit{c}(f, x_i, \hat{y}_i) \cdot \mathit{ea}(\hat{y}_i, Y, \hat{Y})
\end{equation}
where $\mathit{c}(f, x_i, \tilde{y}_i)$ and $\mathit{c}(f, x_i, \hat{y}_i)$ are the model confidence score towards $\tilde{y}_i$ and $\hat{y}_i$. The function $ea$ computes the empirical accuracy of either the model or the user toward their respective label. The empirical accuracy is computed as the cardinality of the intersection between the subset of all their past decisions with label either $\hat{y}_i$ or $\tilde{y}_i$ and the corresponding subset in $Y$, i.e., the final decision, over the subset of all their past decisions with either $\hat{y}_i$ or $\tilde{y}_i$. 
Therefore, each possible label $l \in L$ has two accuracy values -- following the user's and the model's track record. In~\cite{zeni2019fixing}, the user's accuracy values are initialized with $1$, and the model's with $0$ (therefore, the model is not skeptical of earlier decisions).

\smallskip
\textbf{Incremental Decision Tree.}
We employ Extremely Fast Decision Tree (EFDT)~\cite{manapragada2018extremly}, a variant of Hoeffding Tree, which offers performance on par with the non-incremental counterpart~\cite{domingos2000mining, Bifet2009adaptive}. 
EFDT splits a node as soon as the split is deemed useful, with the possibility of later revisiting the decision~\cite{manapragada2018extremly}. 
Being a decision tree, EFDT can also be exploited to provide explanations to the user~\cite{guidotti2018survey}.

\smallskip
\textbf{Preferential Sampling.}
We include an interactive variant of Preferential Sampling (PS), an algorithm increasing group fairness~\cite{kamiran2010classification}.
PS assumes that in the set of class labels $L$ we can recognize a favorable $+$ and an unfavorable $-$ decision, i.e., $L = \{+,-\}$, while among $A$ we can denote a binary sensitive attribute $\mathit{sa} \in A$, e.g., \textit{Sex}. 
The possible values $\{v, \bar{v}\}$ of $\mathit{sa}$ refers to a discriminated group $v$ and privileged group $\bar{v}$, e.g., \textit{Female} and \textit{Male}.
The algorithm identifies the size of the groups of \textit{D}iscriminated records with a \textit{P}ositive (DP) or \textit{N}egative label (DN), and of \textit{P}rivileged records with a \textit{P}ositive (PP) or \textit{N}egative label (PN). 
Given $X$, it computes the \textit{dataset discrimination score} as:
\begin{equation} 
\centering
\label{eq:PS_disc}
  disc(X, \mathit{sa}, v) = \nicefrac{|\mathit{PP}|}{|\mathit{PP}\cup\mathit{PN}|} - \nicefrac{|\mathit{DP|}}{|\mathit{DP}\cup\mathit{DN}|}
\end{equation}
Then, it computes how many records from PP and DN should be removed, and how many from DP and PN should be duplicated to reach $\mathit{disc} \approx 0$. 
Records are selected w.r.t. the prediction probability of a classifier trained on $X$. 
A variant supporting non-binary sensitive attributes, and where the user does not need to know \textit{a priori} the discriminated group(s), 
is presented in~\cite{mazzoni2022genetic}.

\section{A Frank System}
\label{sec:method}
\frank{} is a system for HDM, learning from the decisions of the human decision-maker (typically identified as the ``user''), continuously evolving with them, and aiding the human to remain consistent by offering suggestions and explanations. 
\frank{} is named after its frank behavior -- it interacts with the user as soon as something ``unexpected'' happens.
Other than \frank{} and the user, in line with~\cite{jarrett2022online}, we also suppose a third agent, i.e., the user's \textit{supervisor}.
Depending on the context, the supervisor could be someone enforcing company policies to the user's decisions, e.g., making sure they are not biased by personal beliefs, or someone with higher expertise than the user, e.g., a senior doctor.

\SetKwComment{Comment}{/* }{ */}

\begin{algorithm}[t]
\caption{\frank{}}\label{alg:algoFrank}
    \SetKwInOut{Input}{Input}
    \SetKwInOut{Output}{Output}
    \Input{$X$ - records to label, 
    $R$ - supervisor rule set,
    $\mathit{sa}$ - sensitive attribute,
    $s$ - skepticality thr,
    $k$ - nbr of iter. for GFC,
    $\mathit{stp}$ - stopping condition,
    }
    $X', Y', \tilde{Y}, \hat{Y}, \dddot{Y}, f \gets \mathit{initialize}$;\tcp*[f]{\texttt{\scriptsize sets initialization}}\\
    \While(\tcp*[f]{\texttt{\scriptsize until a stop condition is met}}){$\mathit{stop} \neq \mathit{True}$}{
        $x_i \gets \mathit{receive\_record}(X)$;\tcp*[f]{\texttt{\scriptsize receive a new un-label record}}\\
        $\hat{y}_i \gets \mathit{user\_decision(x_i)}$; \tcp*[f]{\texttt{\scriptsize get user decision}}\\
        $\tilde{y}_i \gets \mathit{f(x_i)}$; \tcp*[f]{\texttt{\scriptsize get model prediction}}\\
        \tikzmark{irc_top}\tikzmark{irc_left}
        \uIf(\tcp*[f]{\texttt{\scriptsize if $x_i$ covered by expert rule}}){$\mathit{ideal\_rule_R(x_i)}$}{
        $\bar{y}_i \gets \mathit{rule\_label_R(x_i)}$;\tcp*[f]{\texttt{\scriptsize get $\bar{y}_i$ from rule}}\\ 
          $y_i \gets \bar{y}_i$;\tcp*[f]{\texttt{\scriptsize $\bar{y}_i$ is compulsorily accepted}}}
        \tikzmark{irc_bottom}\tikzmark{ifc_top}\tikzmark{ifc_left}
        \uElseIf(\tcp*[f]{\texttt{\scriptsize if $x_i$ is similar to past records}}){$\mathit{individual\_fairness_{\mathit{sa}}(x_i, X')}$}{
        $y_p' \gets \mathit{get\_similar\_past\_label(x_i, X', Y')}$;\tcp*[f]{\texttt{\scriptsize get $y_p'$ from past records}}\\ 
        
        \If(\tcp*[f]{\texttt{\scriptsize conflict with a past decision}}){
            $\hat{y}_i \neq y_p'$}{$\langle y_i, Y' \rangle \gets \mathit{solve\_conflict}(x_i, y_p', \hat{y}_i, Y')$;\tcp*[f]{\texttt{\scriptsize solve conflict \& update $Y'$}}
        } \lElse(\tcp*[f]{\texttt{\scriptsize otherwise, user decision $\hat{y_i}$ is accepted}}){$y_i \gets \hat{y}_i$}}
        \tikzmark{ifc_bottom}\tikzmark{skp_top}\tikzmark{skp_left}
        \uElseIf(\tcp*[f]{\texttt{\scriptsize if clash \& skepticism }}){$ \hat{y}_i \neq \tilde{y}_i \wedge \mathit{skept}_s(f, x_i, \tilde{y}_i, \hat{y}_i, \dddot{Y}, \tilde{Y}, \hat{Y})$}{
            \If(\tcp*[f]{\texttt{\scriptsize if an explanation for $\tilde{y}_i$ is desired}}){$\mathit{is\_expl\_desired(x_i, \tilde{y}_i)}$}{
                $e_i \gets \mathit{get\_and\_show\_expl}(x_i, \tilde{y}_i, f, X')$;\tcp*[f]{\texttt{\scriptsize return explanation $e_i$}}\\
            }
            \lIf(\tcp*[f]{\texttt{\scriptsize $\tilde{y}_i$ is accepted}}){$\mathit{accept\_label\_change}(x_i, \tilde{y}_i)$}{
                    $y_i \gets \tilde{y}_i$}
                \lElse(\tcp*[f]{\texttt{\scriptsize $\tilde{y}_i$ is refused}}){$y_i \gets \hat{y}_i$}
        }\tikzmark{skp_bottom}
        \lElse(\tcp*[f]{\texttt{\scriptsize otherwise $\hat{y}_i$ is accepted}}){$y_i \gets \hat{y}_i$}
        $X' \gets X' \cup \{ x_i\}$; $Y' \gets Y' \cup \{y_i\}$;
        $\dddot{Y} \gets \dddot{Y} \cup \{y_i\}$; \tcp*[f]{\texttt{\scriptsize update sets }}\\
        $\hat{Y} \gets \hat{Y} \cup \{\hat{y}_i\}$; $\tilde{Y} \gets \tilde{Y} \cup \{\tilde{y}_i\}$; $f \gets \mathit{update(f, x_i, y_i)}$;\tcp*[f]{\texttt{\scriptsize update sets and model}}\\
        \tikzmark{gfc_top}\tikzmark{gfc_left}
        \If(\tcp*[f]{\texttt{\scriptsize every $k$ records}}){$|Y'| \% k = 0$}{
            $Y', f \gets \mathit{group\_fairness\_check_{\mathit{sa}}(X', Y', f)}$; \tcp*[f]{\texttt{\scriptsize run GFC}}\tikzmark{gfc_bottom}
            \AddNoteA{irc_bottom}{irc_top}{irc_left}{carrotorange}{\textbf{IRC}}
\AddNoteA{ifc_bottom}{ifc_top}{ifc_left}{darkgreen}{\textbf{IFC}}  
\AddNoteA{skp_bottom}{skp_top}{skp_left}{darkblue}{\textbf{SLC}}  
\AddNoteA{gfc_bottom}{gfc_top}{gfc_left}{darkpastelred}{\textbf{GFC}}}}
\end{algorithm}

The pseudocode of \frank{} is reported in Algorithm~\ref{alg:algoFrank}. 
\frank{} requires a set of records to label $X$, which are received one by one, 
a set of rules $R$ provided by the user's supervisor, 
a sensitive attribute $\mathit{sa}$,
a skepticality threshold $s$, 
the number of iterations $k$ after which a group fairness check is performed on the records and decisions analyzed so far, 
and a stopping condition $\mathit{stp}$.
At this stage, we are very general about the stopping condition $\mathit{stp}$ as it might be implemented as reaching a certain number of labeled records, or an accuracy higher than a threshold\footnote{\scriptsize In our experiments, we consider as $\mathit{stp}$ a certain number of instances to be analyzed, leaving for future work the study of measures automatically unveiling when to stop the training.} for $f$.
The initialization of $X', Y', \tilde{Y}, \hat{Y}, \dddot{Y}$ in line 1 can rely on empty sets for a cold start execution, or they might be initialized with records and decisions of previous runs.
We use 
$X'$ to collect the set of records analyzed so far, 
$Y'$ for the set of final hybrid decisions taken on the records in $X'$, 
$\tilde{Y}$ for the set of decisions of \frank{}'s EFDT model $f$ alone,
$\hat{Y}$ for the set of decisions proposed by the user alone, and
$\dddot{Y}$ to store the decisions taken by \frank{} and the user without re-labelling due to fairness corrections.
Also, $f$ might be completely untrained, pre-trained non-interactively on some records, or pre-trained in a past run of \frank{}\footnote{\scriptsize In our experiments, we consider the sets $X', Y', \tilde{Y}, \hat{Y}, \dddot{Y}$ initialized with empty sets and $f$ pre-trained non-interactively on 500 records. Future works will investigate further these aspects.}.
Until the stopping condition $\mathit{stp}$ is met (line 2), \frank{} receives a $x_i$ from $X$ (line 3). 
As in SL~\cite{teso2021interactive}, the user assigns a label $\hat{y}_i$, and \frank{}'s model $f$ a label $\tilde{y}_i$, i.e., the prediction (lines 4 and 5).

With \textit{Ideal Rule Check} (IRC), \frank{} checks if the record $x_i$ is covered by a rule in the rule set $R$ provided by the user's supervisor (line 7).
If it is, then the decision $\bar{y}_i$ is derived from the rule and assigned to the final decision $y_i$ (line 8).
If none of the rules from $R$ cover the record, with \textit{Individual Fairness Check} (IFC), \frank{} checks if the user's decision complies with the individual fairness condition by comparing $\hat{y}_i$ to the labels assigned to ``similar'' past records (lines 9-13).
The definition of similarity is further defined below.
\textit{Skeptical Learning Check} (SLC) is triggered if no similar records exist and the user's decision $\hat{y}_i$ and \frank{}'s prediction $\tilde{y}_i$ are not the same. 
If \frank{} is skeptical of $\hat{y}_i$, the user is asked if they want an explanation for $\tilde{y}_i$ (line 15).
If the user accepts, they are shown the explanation $e_i$ (line 16). 
Regardless, the user is then asked if they accept $\tilde{y}_i$ as the final decision $y_i$ (lines 17).
If the user refuses (line 18), if \frank{} is not skeptical, or if it agrees with the user (line 19), the user's decision $\hat{y}_i$ is accepted as the final decision $y_i$.
Regardless of the triggered checks, $x_i$ and $y_i$ are added respectively to $X'$ and $Y'$ (line 20), and are used to update \frank{}'s model $f$ (line 21). 
Similarly, $\tilde{y}_i$ and $\hat{y}_i$ are added to $\tilde{Y}$ and $\hat{Y}$, respectively.
Also, $y_i$ is added to $\dddot{Y}$, which might differ from the set of labels $Y'$ in the case of relabeling. 
Finally, every $k$ records, \frank{} performs \textit{Group Fairness Check} (GFC, lines 24-25), asking the user if they want to change the label of some past records to reduce the dataset's discrimination as computed by Prefential Sampling~\cite{kamiran2010classification}. 
\frank{} prioritizes IRC to follow the guidelines of the supervisor, then IFC for fairness among similar records, 
and, finally, SLC. 
To avoid contradictions, once a final label $y_i$ is set, checks with less priority are never triggered, and GFC cannot relabel records labeled by IRC or IFC. 
We stress that the user \textit{has} to accommodate suggestions offered by IRC and IFC. 
On the other hand, the user is free to disregard suggestions by SLC and GFC.
Depending on the use cases, certain checks might be turned off, e.g., IFC and GFC in health contexts. 
As some functions cycle the previously-seen records, \frank{}'s algorithmic complexity is $O(n)$ with $n = |X'|$.

In the following, we provide a detailed explanation of \frank{}'s four checks.

\smallskip
\textbf{Ideal Rule Check.}
Each rule $r \in R$ includes a set of conditions and a label $\bar{y}$. 
The $\mathit{ideal\_rule}$ function checks if $x_i$ follows the conditions of one of the rules in $R$ (line 6), and if it is, it provides the label $\bar{y}_i$ (line 7), which is selected as the final decision $y_i$, regardless of the user's label $\hat{y}_i$. 
In case of divergence between the user's decision and the supervisor's rule, the user is notified that their decision is not compliant. 
Since IRC leaves no freedom of choice, the rules $R$ should only cover very limited, specific, and ideal cases, describing records which should \textit{absolutely} receive a certain label. 
The supervisor should also make sure the rules $R$ are mutually exclusive. 
Besides, to avoid conflicts with fairness-related functions, the rules' conditions should not rely on sensitive attributes.

\smallskip
\textbf{Individual Fairness Check.}
IFC is meant to reduce the pairs of records violating individual fairness condition, i.e., similar individuals should be treated similarly, by assessing if records similar to $x_i$ received a different label than $\hat{y_i}$. 
\frank{} checks through the $\mathit{individual\_fairness}$ function (line 9) if there is at least one past record $x_p' \in X'$ identical or ``similar'' to the current record $x_i$.
Given a binary sensitive attribute $\mathit{sa} \in A$, \frank{} defines two records $x_i$ and $x_p'$ \textit{similar} if $v_j = v'_j \forall a_j \in A - \{\mathit{sa}\}$, i.e., $x_i$ and $x_p'$ are similar if they are identical, save for the value of $\mathit{sa}$.
More than one similar or identical record $x'_p \in X$ can be found, and, by construction, they have all the same past label $y_p' \in Y'$ (line 10).
If there is a disagreement between the current decision and past decisions, i.e., $y_p' \neq \hat{y_i}$ (line 11), then in line 12 $\mathit{solve\_conflict}$ prompts the user either to change their decision to make it compliant with past records, i.e., to select $y_p'$ as $y_i$, or to keep the decision but relabel past records with $\hat{y_i}$, i.e., modifying the labels in $Y'$\footnote{\scriptsize Note that $\dddot{Y}$ is not modified, nor taken into account by IFC.}.
If the latter is chosen, $f$ is also retrained, accounting for the modified labels.
Otherwise, if $y_p' = \hat{y_i}$, $x_i$ is assigned $\hat{y_i}$, i.e., the user's decision is accepted (line 13) as it is consistent with past records.

\smallskip
\textbf{Skeptical Learning Check.}
If there is a disagreement between the decision of the user and $f$, i.e., $\hat{y}_i \neq \tilde{y}_i$,
the $\mathit{skept}$ function (line 14) computes \frank{}'s \textit{skepticality} following Eq.~\ref{eq:skept}. 
If it is higher than $s$, \frank{} is skeptical. 
Empirical accuracy values are initialized as in Sec.~\ref{sec:background}.
We emphasize that $\mathit{skept}$ does not take as input $Y'$, i.e., the set of decisions after possible re-labeling, but $\dddot{Y}$, i.e., the set of decisions made by the user after \frank{}'s checks for each record\footnote{\scriptsize $Y'$ and $\dddot{Y}$ coincide until the user relabel older records if prompted by IFC or GFC.}.
If skeptical, \frank{} proposes $\tilde{y}_i$ for $y_i$, and asks the user if they want an explanation $e_i$ (line 15).
The user is then asked to accept $\tilde{y}_i$ (line 17).
The user has the full veto power against \frank{}, and if they reject $\tilde{y}_i$, the user label is collected as the final decision $y_i$ (line 18).
If the user accepts to see an explanation, \frank{} runs the $\mathit{get\_and\_show\_expl}$ function and provides it to the user (line 16).
\frank{} can provide \textit{Logic-based Explanations}, where a global representation of the EFDT is shown alongside the local decision rule followed for the record $x_i$ and $\tilde{y}_i$ (such as \textit{IF Years\_of\_Experience $>$ 5 AND Attitude $=$ True THEN Hire}), or
\textit{Instance-based Explanations}, i.e., records similar to $x_i$ which can be either \textit{real} or \textit{synthetic}. 
These records are classified by $f$ either with $\tilde{y}_i$, i.e., an \textit{example} of \frank{}'s decision, or $\hat{y}_i$, i.e., a \textit{counter-example}.
\frank{}'s explanations are the result of a \textit{co-evolutionary relationship} with the user, leading to more detailed justifications over time.
Thus, the user should progressively trust \frank{} more.

\smallskip
\textbf{Group Fairness Check.} 
GFC checks if one of the value of a binary sensitive attribute $\mathit{sa} \in A$ are discriminating against the other group w.r.t. $Y'$. 
GFC is independent from the other checks, and it is always triggered every $k$ records (see lines 22-23).
\frank{} computes $\mathit{disc}$ and the DN, DP, PN, and PP groups of the set of records $X'$ w.r.t. the labels $Y'$, following~\cite{mazzoni2022genetic}.
Then, it orders the records from DN and PP following the prediction probability of $f$, and calculates how many of them should be removed. 
Finally, the records with higher probability are shown to the user, who can choose to change their label. 
The new labels replace the older ones in $Y'$, and $f$ is retrained from scratch.
Thus, GFC is an interactive implementation of PS, where the user is made aware of their discriminating behavior and is asked to relabel past records to mitigate the discrimination. 

\section{Experiments}
\label{sec:experiments}
We evaluated \frank{}\footnote{\scriptsize The Python code is available here: \url{https://github.com/FedericoMz/Frank}.} on three real-world datasets and, in line with~\cite{bontempelli2020learning, joshi2021learning}, we employed simulated users to assess its impact in a variety of conditions.

\smallskip
\textbf{Users.}
We employed five kinds of \textit{simulated} users:
the \textit{Real Expert}, who always makes decisions following the ground truth (which is unknown in a real scenario), 
the \textit{Absent-Minded}, an easily-distracted expert who follows the ground truth 75\% of times, 
the \textit{Coin-Tosser}, who makes decisions by flipping a coin, 
and the \textit{Bayesian} and \textit{Similarity} experts, simulated by Naive Bayes and KNN~\cite{tan2016introduction}. 
For IFC, we suppose that all the users have conservative behavior w.r.t. their past decisions, with 80.00\% of chance of changing the label assigned to the current record $x_i$, instead of re-labeling past records. 
For SLC, we set a threshold $s$ of $0.05$, increasing the times \frank{} is skeptical. We assumed that the users can always \textit{accept} or \textit{decline} \frank{}'s suggestions, or \textit{randomly} choose.
For Bayesian and Similarity experts, we also envisioned users who request \textit{explanations}, i.e., five synthetic examples and counterexamples, monitoring their reaction\footnote{\scriptsize As synthetic records lack a ground truth, this option cannot be implemented with the other users.}. 
If they agree with more than half, they accept \frank{}'s suggestions. 
For GFC, we suppose that the user selects to re-label the top 25\% DN and PP records.

\smallskip
\textbf{Datasets.}
The \texttt{Adult}, \texttt{COMPAS} and \texttt{HR} datasets\footnote{\scriptsize 
\url{kaggle.com/datasets/}.
} simulate classification tasks for granting credits, predicting recidivism, or giving a promotion, i.e., possible real use-cases for \frank{}. 
In \texttt{HR}, only $8\%$ of records belong to the positive class, compared to the $25\%$ and the $50.00\%$ in \texttt{Adult} and \texttt{COMPAS}, which are, however, highly discriminating \cite{quy2021datasets}.
In contrast, \texttt{HR} is fair w.r.t. \textit{Sex}.
After removing duplicated or incomplete records,
we randomly selected 2,000 records to incrementally train \frank{}, i.e., $X$. We set labeling all the records in $X$ as our stopping condition $stp$.
The Naive Bayes and KNN models were trained on an additional 500 records.
Half of them were also used to pre-train \frank{}'s ML model $f$.
Finally, a dataset $X_T$ includes 500 records reserved to test $f$.
For IRC, we set the following rules: for \texttt{Adult}, IF $\mathit{capital\_gain} > 9000$ THEN $\bar{y} = +$; for \texttt{COMPAS}, IF $\mathit{priors\_count} > 0$ THEN $\bar{y} = +$; for \texttt{HR}, IF $\mathit{awards\_won} = \mathit{True} $ THEN $\bar{y} = +$.

\smallskip
\textbf{Evaluation Measures.}
We measured the \textit{Co-evolutionary Accuracy} (\textit{CA}) by comparing $Y'$ with the ground truth $Y$, and the Model Accuracy (\textit{MA}) by comparing the prediction of $f$ on $X_T$ with its ground truth $Y_T$.
Likewise, we measured the \textit{Co-evolutionary Discrimination} (\textit{CD}) and the Model Discrimination (\textit{MD}). 
The $\mathit{disc}$ score was computed towards \textit{Female} for all datasets\footnote{\scriptsize Note that a negative $\mathit{disc}$ implies that \textit{Male} is discriminated.}. 
Finally, we counted the number of Unfair Couples (\textit{UC}), i.e., similar records violating individual fairness by receiving a different label.
Ideal values are $1$ for \textit{CA} and \textit{MA}, $0$ for the others. 
Each experiment was repeated 10 times. 
The tables report the average results, standard deviations are very low and not reported.

\begin{table}[t]
\centering

\scriptsize
\setlength{\tabcolsep}{0.5mm}
\caption{Ablation study of \frank{}'s checks.}
\label{tab:early_results}

    \begin{tabular}{cc|ccccc|ccccc|ccccc|ccccc}
    & & \multicolumn{5}{c}{\textit{None}} & \multicolumn{5}{c}{\textit{oIRC}} & \multicolumn{5}{c}{\textit{oIFC}} & \multicolumn{5}{c}{\textit{oGFC}} \\
    & &
     CA & MA & CD & MD & UC & CA & MA & CD & MD & UC & CA & MA & CD & MD & UC & CA & MA & CD & MD & UC \\
    \hline
   \multirow{5}{*}{\rotatebox[origin=c]{90}{\texttt{Adult}}} & \textit{Real} 
    & 1.0 & .83 & .21 & .18 & 7.0 &
.96 & .82 & .23 & .15 & 7.0 &
1.0 & .84 & .22 & .17 & 0.0 &
.84 & .75 & -.02 & .01 & 6.0 \\
    & \textit{Abs.} & .75 & .77 & .10 & .09 & 5.3 &
.74 & .76 & .13 & .09 & 5.3 &
.75 & .77 & .11 & .12 & 0.0 &
.78 & .76 & .01 & .04 & 4.2 \\
    & \textit{Coin} & .50 & .56 & .00 & .02 & 5.6 &
.52 & .51 & .03 & -.01 & 5.6 &
.50 & .52 & .00 & .00 & 0.0 &
.55 & .55 & .03 & .04 & 5.3 \\
    & \textit{Bayes} & .80 & .77 & .12 & .07 & 0.0 &
.79 & .76 & .11 & .09 & 0.0 &
.80 & .77 & .12 & .07 & 0.0 &
.80 & .77 & .09 & .09 & 0.0 \\
    & \textit{Sim.} & .79 & .76 & .20 & .24 & 1.0 &
.79 & .76 & .20 & .24 & 1.0 &
.79 & .76 & .20 & .24 & 0.0 &
.80 & .77 & .03 & .17 & 0.0 \\
    \hline

    \multirow{5}{*}{\rotatebox[origin=c]{90}{\texttt{COMPAS}}} & \textit{Real} & 
    1.0 & .69 & -.14 & -.21 & 42. &
.65 & .61 & -.15 & -.21 & 18. &
.98 & .68 & -.14 & -.24 & 0.0 &
.75 & .64 & -.06 & -.15 & 17. \\
    & \textit{Abs.} & .75 & .63 & -.07 & -.19 & 50. &
.60 & .61 & -.12 & -.21 & 24. &
.74 & .64 & -.08 & -.19 & 0.0 &
.64 & .62 & -.03 & -.17 & 24. \\
    & \textit{Coin} & .50 & .57 & .00 & -.17 & 56. &
.54 & .55 & -.09 & -.08 & 27. &
.50 & .52 & -0.0 & -.09 & 0.0 &
.49 & .48 & .01 & -.01 & 32. \\
    & \textit{Bayes} & .63 & .63 & -.20 & -.19 & 0.0 &
.59 & .62 & -.18 & -.25 & 0.0 &
.63 & .63 & -.20 & -.19 & 0.0 &
.61 & .63 & -.15 & -.18 & 0.0 \\
    & \textit{Sim.} & .63 & .66 & -.31 & -.18 & 25. &
.58 & .62 & -.21 & -.25 & 15. &
.62 & .66 & -.28 & -.17 & 0.0 &
.63 & .66 & -.01 & -.21 & 17. \\
    \hline
    \multirow{5}{*}{\rotatebox[origin=c]{90}{\texttt{HR}}} & \textit{Real} & 
    1.0 & .93 & -.01 & .00 & 39. &
.99 & .89 & -.02 & 0.02 & 39. &
.98 & .93 & -.01 & .00 & .99 &
.94 & .93 & .00 & .00 & 31. \\
    & \textit{Abs.} & .75 & .93 & -.01 & .00 & 24. &
.74 & .92 & -.01 & -0.01 & 24. &
.74 & .93 & .00 & .00 & 0.0 &
.85 & .93 & .00 & .00 & 20. \\
    & \textit{Coin} & .50 & .93 & -.01 & .00 & 21. &
.50 & .83 & -.02 & -.06 & 21. &
.50 & .93 & .01 & .00 & 0.0 &
.62 & .65 & .01 & .06 & 19. \\
    & \textit{Bayes} & .89 & .92 & .00 & -.02 & 0.0 &
.89 & .92 & .00 & -.02 & 0.0 &
.89 & .92 & .00 & -.02 & 0.0 &
.90 & .93 & .00 & .00 & 0.0 \\
    & \textit{Sim.} & .89 & .93 & .00 & .00 & 0.0 &
.89 & .91 & .00 & -.02 & 0.0 &
.89 & .93 & .00 & .00 & 0.0 &
.89 & .93 & .00 & .00 & 0.0 \\
    \hline
    \end{tabular}
\end{table}

\smallskip
\textbf{Results.}
As an ablation study of \frank{}'s structure, in Table~\ref{tab:early_results}, we report the results when \textit{None} of \frank{}'s functions are enabled, and when only IRC, IFC, or GFC are enabled (\textit{oIRC}, \textit{oIFC}, \textit{oGFC}).
The impact of IRC is minimal on \texttt{Adult} and \texttt{HR}, whereas it negatively affects all the experts except for the \textit{Coin-Tosser} in \texttt{COMPAS}. This is probably due to the selected rules, either too narrow in scope or inaccurate. These results highlight the importance of selecting good rules for \frank{}.
On the other hand, comparing the \textit{oIFC} and \textit{oGFC} columns to \textit{None}, we can see a significant improvement in terms of fairness. 
IFC always successfully minimizes \textit{UC} with no side effects, whereas GFC consistently reduces both \textit{CD} and \textit{MD}.
For \texttt{Adult} and \texttt{COMPAS} and with the \textit{Real Expert}, this is at the expense of \textit{CA} and \textit{MA}. 
However, we should stress that the ``accuracy'' of very biased datasets does not necessarily mirror ``right'' decisions. 
In fact, on the already balanced \texttt{HR}, the impact on \textit{CA} and \textit{MA} with the \textit{Real Expert} is minimal. Additionally, with \texttt{Adult} and \texttt{HR}, GFC improves the accuracy of \textit{Absent-Minded} and \textit{Coin-Tosser} experts without negatively impacting the model-based ones.

\begin{table}[!t]
    \centering
    \caption{\frank{} vs traditional SL. Best scorer in bold, parity in italics.}
    \scriptsize
    \setlength{\tabcolsep}{1.5mm}
    \centering
    \begin{tabular}{ccc|cc|cc|cc|cc|cc}
    \multirow{2}{*}{ } & \multirow{2}{*}{} &  & \multicolumn{2}{c}{\textit{Real Expert}}  & \multicolumn{2}{c}{\textit{Absent-Minded}}  & \multicolumn{2}{c}{\textit{Coin-Tosser}}  & \multicolumn{2}{c}{\textit{Bayesian}}  &
    \multicolumn{2}{c}{\textit{Similarity}}  \\ 
    & & & \textsc{SL} & \frank{} & \textsc{SL} & \frank{} & \textsc{SL} & \frank{} & \textsc{SL} & \frank{} & \textsc{SL} & \frank{}\\
    
    \hline 
    \multirow{15}{*}{\rotatebox[origin=c]{90}{\texttt{Adult}}} &

    \multirow{5}{*}{\rotatebox[origin=c]{90}{\textit{accept}}}
    
     & CA &
0.74 & \textbf{0.78} & 
0.73 & \textbf{0.78} & 
0.73 & \textbf{0.78} & 
0.73 & \textbf{0.78} & 
0.74 & \textbf{0.77} \\ 
& & MA &
0.66 & \textbf{0.75} & 
0.66 & \textbf{0.75} & 
0.65 & \textbf{0.75} & 
0.64 & \textbf{0.75} & 
0.74 & \textbf{0.75} \\ 
& & CD &
\textbf{0.02} & 0.05 & 
\textbf{0.03} & 0.05 & 
\textbf{0.04} & 0.05 & 
\textbf{0.04} & 0.05 & 
\textbf{0.00} & 0.01 \\ 
& & MD &
-0.10 & \textbf{0.05} & 
-0.06 & \textbf{0.05} & 
\textbf{-0.01} & 0.05 & 
\textbf{0.00} & 0.05 & 
-0.02 & \textbf{0.01} \\ 
& & UC &
1.00 & \textbf{0.00} & 
1.00 & \textbf{0.00} & 
1.00 & \textbf{0.00} & 
1.00 & \textbf{0.00} & 
\textit{0.00} & \textit{0.00} \\

    \cline{2-13}
    & \multirow{5}{*}{\rotatebox[origin=c]{90}{\textit{decline}}}

     & CA &
\textbf{1.00} & 0.83 & 
0.75 & \textbf{0.77} & 
0.50 & \textbf{0.57} & 
\textbf{0.80} & 0.79 & 
0.79 & \textbf{0.80} \\ 
& & MA &
\textbf{0.83} & 0.75 & 
\textbf{0.77} & 0.75 & 
0.56 & \textbf{0.58} & 
\textbf{0.77} & 0.76 & 
0.76 & \textbf{0.77} \\ 
& & CD &
0.21 & \textbf{0.03} & 
0.11 & \textbf{0.01} & 
\textit{-0.01} & \textit{-0.01} & 
0.12 & \textbf{0.11} & 
0.20 & \textbf{0.03} \\ 
& & MD &
0.18 & \textbf{0.05} & 
0.12 & \textbf{0.05} & 
\textbf{-0.04} & 0.05 & 
\textbf{0.07} & 0.09 & 
0.24 & \textbf{0.17} \\ 
& & UC &
7.00 & \textbf{0.00} & 
6.10 & \textbf{0.00} & 
5.60 & \textbf{0.00} & 
0.00 & \textbf{0.00} & 
1.00 & \textbf{0.00} \\

    \cline{2-13}
    & \multirow{5}{*}{\rotatebox[origin=c]{90}{\textit{random}}}

    & CA &
\textbf{0.89} & 0.80 & 
0.74 & \textbf{0.77} & 
0.55 & \textbf{0.57} & 
\textit{0.79} & \textit{0.79} & 
\textit{0.77} & \textit{0.77} \\ 
& & MA &
\textbf{0.76} & 0.75 & 
0.71 & \textbf{0.75} & 
\textit{0.58} & \textit{0.58} & 
\textit{0.76} & \textit{0.76} & 
0.73 & \textbf{0.75} \\ 
& & CD &
0.14 & \textbf{0.03} & 
0.09 & \textbf{0.01} & 
0.04 & \textbf{0.01} & 
0.10 & \textbf{0.08} & 
0.14 & \textbf{0.00} \\ 
& & MD &
0.09 & \textbf{0.05} & 
\textbf{0.04} & 0.05 & 
0.09 & \textbf{-0.01} & 
\textbf{0.07} & 0.08 & 
0.14 & \textbf{0.02} \\ 
& & UC &
4.60 & \textbf{0.00} & 
3.10 & \textbf{0.00} & 
4.20 & \textbf{0.00} & 
0.10 & \textbf{0.00} & 
0.60 & \textbf{0.00} \\

    \hline

    \multirow{15}{*}{\rotatebox[origin=c]{90}{\texttt{COMPAS}}} &
    
    \multirow{5}{*}{\rotatebox[origin=c]{90}{\textit{accept}}}

    & CA &
\textit{0.52} & \textit{0.52} & 
0.51 & \textbf{0.53} & 
0.52 & \textbf{0.54} & 
\textit{0.52} & \textit{0.52} & 
0.55 & \textbf{0.58} \\ 
& & MA &
\textbf{0.52} & 0.51 & 
0.49 & \textbf{0.54} & 
0.49 & \textbf{0.55} & 
\textbf{0.52} & 0.51 & 
0.56 & \textbf{0.62} \\ 
& & CD &
-0.02 & \textbf{0.00} & 
-0.01 & \textbf{0.00} & 
-0.01 & \textbf{0.00} & 
-0.02 & \textbf{0.00} & 
-0.09 & \textbf{-0.05} \\ 
& & MD &
\textbf{-0.02} & -0.04 & 
\textbf{0.01} & -0.07 & 
\textbf{0.01} & -0.08 & 
\textbf{-0.02} & -0.04 & 
\textbf{-0.03} & -0.11 \\ 
& & UC &
8.00 & \textbf{0.00} & 
17.60 & \textbf{0.00} & 
17.60 & \textbf{0.00} & 
8.00 & \textbf{0.00} & 
1.00 & \textbf{0.00} \\

    \cline{2-13}
    & \multirow{5}{*}{\rotatebox[origin=c]{90}{\textit{decline}}}

    & CA &
\textbf{1.00} & 0.77 & 
\textbf{0.75} & 0.64 & 
\textit{0.50} & \textit{0.50} & 
\textbf{0.63} & 0.61 & 
\textbf{0.63} & 0.62 \\ 
& & MA &
\textbf{0.69} & 0.66 & 
\textbf{0.65} & 0.62 & 
\textbf{0.53} & 0.52 & 
\textit{0.63} & \textit{0.63} & 
\textbf{0.66} & 0.65 \\ 
& & CD &
-0.14 & \textbf{0.00} & 
-0.06 & \textbf{-0.01} & 
0.01 & \textbf{0.00} & 
-0.20 & \textbf{-0.15} & 
-0.31 & \textbf{-0.04} \\ 
& & MD &
-0.21 & \textbf{-0.19} & 
-0.21 & \textbf{-0.13} & 
\textbf{-0.07} & -0.10 & 
-0.19 & \textbf{-0.18} & 
\textit{-0.18} & \textit{-0.18} \\ 
& & UC &
42.00 & \textbf{0.00} & 
50.00 & \textbf{0.00} & 
51.10 & \textbf{0.00} & 
\textit{0.00} & \textit{0.00} & 
25.00 & \textbf{0.00} \\

    \cline{2-13}
    & \multirow{5}{*}{\rotatebox[origin=c]{90}{\textit{random}}}

    & CA &
\textbf{0.80} & 0.66 & 
\textbf{0.65} & 0.58 & 
0.50 & \textbf{0.54} & 
\textbf{0.62} & 0.60 & 
\textit{0.62} & \textit{0.62} \\ 
& & MA &
\textbf{0.64} & 0.61 & 
\textbf{0.57} & 0.56 & 
0.48 & \textbf{0.57} & 
\textbf{0.63} & 0.60 & 
0.65 & 0.65 \\ 
& & CD &
-0.15 & \textbf{-0.01} & 
-0.10 & \textbf{0.00} & 
-0.02 & \textbf{-0.01} & 
-0.18 & \textbf{-0.08} & 
-0.23 & \textbf{-0.05 }\\ 
& & MD &
-0.16 & \textbf{-0.12} & 
-0.17 & \textbf{-0.11} & 
\textbf{0.00} & -0.08 & 
-0.18 & \textbf{-0.15} & 
-0.17 & \textbf{-0.16} \\ 
& & UC &
34.00 & \textbf{0.00} & 
45.20 & \textbf{0.00} & 
43.30 & \textbf{0.00} & 
3.20 & \textbf{0.00} & 
23.70 & \textbf{0.00} \\

   \hline 
    \multirow{15}{*}{\rotatebox[origin=c]{90}{\texttt{HR}}} &
    
    \multirow{5}{*}{\rotatebox[origin=c]{90}{\textit{accept}}}

    & CA &
\textbf{0.90} & 0.89 & 
\textbf{0.90} & 0.86 & 
\textbf{0.90} & 0.88 & 
\textbf{0.90} & 0.89 & 
\textbf{0.90} & 0.89 \\ 
& & MA &
\textbf{0.93} & 0.92 & 
\textbf{0.93} & 0.89 & 
\textbf{0.93} & 0.91 & 
\textbf{0.93} & 0.92 & 
\textbf{0.93} & 0.92 \\ 
& & CD &
\textit{0.00} & 0.01 & 
\textit{0.00} & 0.01 & 
\textit{0.00} & \textit{0.00} & 
\textit{0.00} & \textit{0.00} & 
\textit{0.00} & \textit{0.00} \\ 
& & MD &
\textbf{0.00} & -0.02 & 
\textbf{0.00} & -0.02 & 
\textbf{0.00} & -0.02 & 
\textbf{0.00} & -0.02 & 
\textbf{0.00} & -0.02 \\ 
& & UC &
\textit{0.00} & \textit{0.00} & 
\textit{0.00} & \textit{0.00} & 
\textit{0.00} & \textit{0.00} & 
\textit{0.00} & \textit{0.00} & 
\textit{0.00} & \textit{0.00} \\

    \cline{2-13}
    & \multirow{5}{*}{\rotatebox[origin=c]{90}{\textit{decline}}}

    & CA &
\textbf{1.00} & 0.91 & 
0.75 & \textbf{0.83} & 
0.50 & \textbf{0.66} & 
\textit{0.89} & \textit{0.89} & 
\textit{0.89} & \textit{0.89} \\ 
& & MA &
\textbf{0.93} & 0.92 & 
\textbf{0.93} & 0.92 & 
\textbf{0.90} & 0.71 & 
\textit{0.92} & \textit{0.92} & 
\textbf{0.93} & 0.92 \\ 
& & CD &
\textit{-0.01} & \textit{-0.01} & 
-0.01 & \textbf{0.00} & 
\textit{-0.01} & \textit{-0.01} & 
\textit{0.00} & \textit{0.00} & 
\textit{0.00} & \textit{0.00} \\ 
& & MD &
\textbf{0.00} & -0.02 & 
-0.01 & \textbf{0.00} & 
\textbf{-0.01} & 0.05 & 
\textit{-0.02} & \textit{-0.02} & 
\textbf{0.00} & -0.02 \\ 
& & UC &
39.00 & \textbf{0.00} & 
26.9 & \textbf{0.00} & 
22.6 & \textbf{0.00} & 
\textit{0.00} & \textit{0.00} & 
\textit{0.00} & \textit{0.00} \\

    \cline{2-13}
    & \multirow{5}{*}{\rotatebox[origin=c]{90}{\textit{random}}}

    & CA &
\textbf{0.95} & 0.91 & 
0.82 & \textbf{0.86} & 
0.70 & \textbf{0.82} & 
\textit{0.89} & \textit{0.89} & 
\textit{0.89} & \textit{0.89} \\ 
& & MA &
\textbf{0.93} & 0.92 & 
\textbf{0.93} & 0.90 & 
\textbf{0.93} & 0.86 & 
\textbf{0.93} & 0.92 & 
\textbf{0.93} & 0.92 \\ 
& & CD &
-0.01 & \textbf{0.00} & 
\textit{0.00} & \textit{0.00} & 
\textbf{0.00} & 0.01 & 
\textit{0.00} & \textit{0.00} & 
\textit{0.00} & \textit{0.00} \\ 
& & MD &
\textbf{0.00} & -0.02 & 
\textbf{0.00} & -0.02 & 
\textbf{0.00} & 0.03 & 
\textbf{0.00} & -0.02 & 
\textbf{0.00} & -0.02 \\ 
& & UC &
18.10 & \textbf{0.00} & 
17.40 & \textbf{0.00} & 
16.20 & \textbf{0.00} & 
\textit{0.00} & \textit{0.00} & 
\textit{0.00} & \textit{0.00} \\ 

   \hline 
   
    \end{tabular}
\label{tab:final_results}
\end{table}

Table~\ref{tab:final_results} compares traditional \textsc{SL}~\cite{teso2021interactive} with \frank{} with everything enabled, except for IRC in \texttt{COMPAS}. 
As mentioned for IRC, \frank{} consistently minimizes UC.
In \texttt{Adult}, \frank{} provides each expert better CA and MA if they always accept the suggestions, whereas CD and sometimes MD is slightly better with \textsc{SL}. 
By declining the suggestions or randomizing the choices with \textsc{SL}, the \textit{Real Expert} gets better CA and MA, but worse CD and MD. 
With other experts, \frank{} is better than, or very close to, \textsc{SL} for CA and MA, while consistently improving CD and MD. 
In \texttt{COMPAS}, \frank{} always has a better CD, and often a better MD. 
When the \textit{Real Expert} and the \textit{Absent-Minded} randomize or decline, this is at the expense of CA and, to a lesser extent, MA, with a strong fairness-performance trade-off. 
In other cases, \frank{} performs a bit better or on par with \textsc{SL}.
As for \texttt{HR}, the two methods are very close for the \textit{Real}, \textit{Bayesian}, and \textit{Similarity} experts, with \textsc{SL} slightly better. 
With the \textit{Absent-Minded} and the \textit{Coin-Tosser}, declining or randomizing decision greatly enhances the CA. 
In fact, the randomizing \textit{Coin-Tosser} reaches a CA comparable to the \textit{Absent-Minded}'s. 
Also, with the same example we can notice a lower MA than \textsc{SL}'s. 
This might be due to the fact that IRC, IFC, and GFC are not triggered when $f$ makes decisions on $X_T$.

 \begin{figure}[t]
     \centering
     \includegraphics[trim={0 0.9cm 0 0},clip,width=0.33\textwidth]{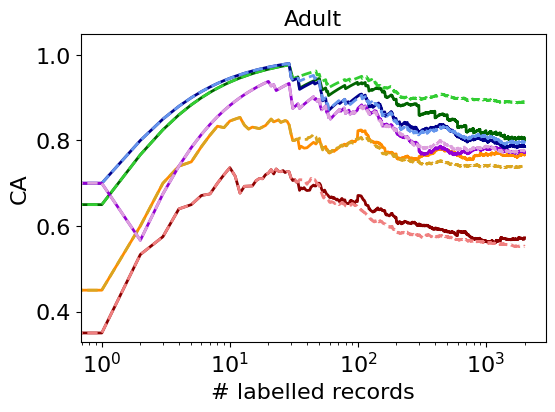}%
     \includegraphics[trim={0 0.9cm 0 0},clip,width=0.33\textwidth]{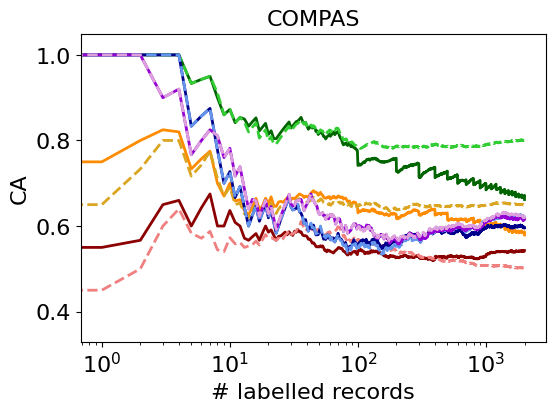}%
     \includegraphics[trim={0 0.9cm 0 0},clip,width=0.33\textwidth]{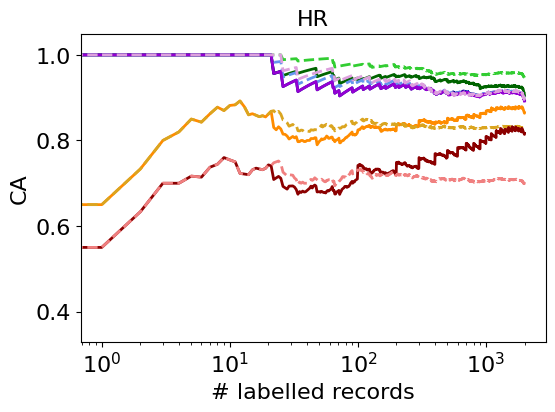} 
     
     \includegraphics[trim={0 0 0 0 cm},clip,width=0.33\textwidth]{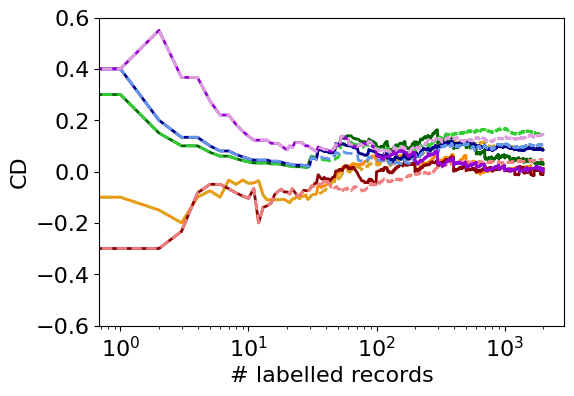}%
     \includegraphics[trim={0 0 0 0 cm},clip,width=0.33\textwidth]{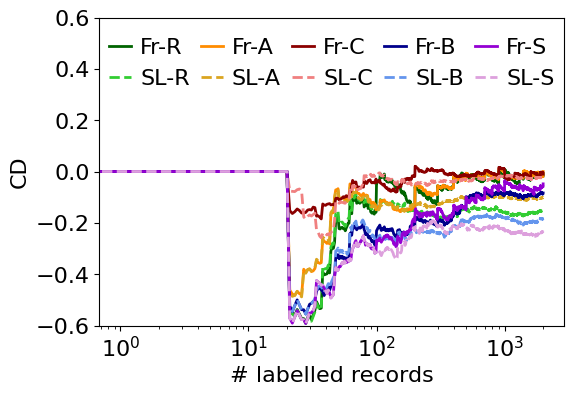}%
     \includegraphics[trim={0 0 0 0 cm},clip,width=0.33\textwidth]{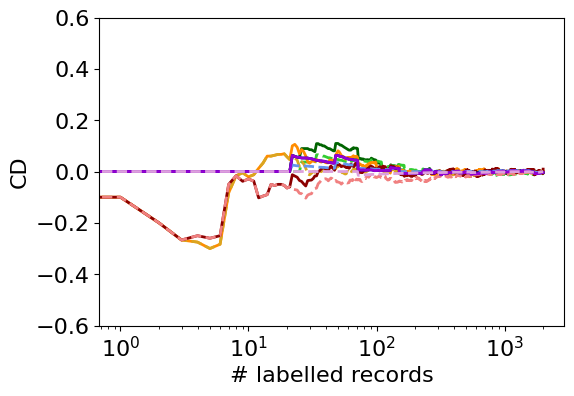} 
     \caption{CA and CD evolution over time with different experts.} 
     \label{fig:evolution}
 \end{figure}
 
Figure~\ref{fig:evolution} shows CD and CA over time for different experts, randomly accepting the suggestions from \frank{} and \textsc{SL}. 
Plots are in log scale along the x-axis.
At first, for each user \frank{} and \textsc{SL} follow a similar pattern, both in terms of CA and CD.
Their lines then diverge due to fairness interventions. 
In \texttt{Adult}, this results in a drop of CA for the \textit{Real Expert}, and in \texttt{COMPAS} also for the \textit{Absent-Minded}. 
In \texttt{HR}, the \textit{Real Expert} is far less affected, as the dataset is less biased. 
On the other hand, the \textit{Absent-Minded} and \textit{Coin-Tosser} receive a noticeable boost in terms of CA. 
In \texttt{Adult} and \texttt{COMPAS}, the \textit{Real} and the \textit{Similarity} experts make biased decisions while paired with \textsc{SL}, whereas their CD with \frank{} is near $0$. \frank{}'s CD lines tend to converge to $0$ for all the datasets. 

\begin{table}[t]
    \centering
    \caption{Users accepting suggestions randomly (RND) or w.r.t. explanations (XAI).}
    \scriptsize
    \setlength{\tabcolsep}{1.2mm}
    \centering
    \begin{tabular}{r|cc|cc|cc|cc|cc|cc}
 & \multicolumn{4}{c}{\texttt{Adult}}
& \multicolumn{4}{c}{\texttt{COMPAS}}
& \multicolumn{4}{c}{\texttt{HR}}
\\
    
    & \multicolumn{2}{c}{\textit{Bayesian}}  & \multicolumn{2}{c|}{\textit{Similarity}}
    & \multicolumn{2}{c}{\textit{Bayesian}}  & \multicolumn{2}{c|}{\textit{Similarity}}
    & \multicolumn{2}{c}{\textit{Bayesian}}  & \multicolumn{2}{c}{\textit{Similarity}}

    \\ 
    & 
    RND & XAI & 
    RND & XAI & 
    RND & XAI & 
    RND & XAI & 
    RND & XAI & 
    RND & XAI \\
    
    \hline 

Agr \% & 96.38 & 88.72 & 
77.14 & 77.37 & 89.64 & 
74.53 & 
68.83 & 60.42 & 100.00 & 100.00 & 99.12 & 99.16 \\ 
Ske \% & 3.49 & 11.20 & 22.47 & 22.47 & 10.32 & 25.41 & 31.04 & 39.45 & 0.00 & 0.00 & 0.79 & 0.73 \\ 
Dis \% & 0.11 & 0.05 & 0.37 & 0.15 & 
0.03 & 0.05 & 0.12 & 0.13 & 0.00 & 0.00 & 0.08 & 0.10 \\     
\hline
Acc \% & 51.99 & 94.03 & 49.61 & 37.58 & 50.49 & 93.94 & 50.37 & 74.02 & N/A & N/A & 54.30 & 0.00 \\ 
Dec \% & 48.01 & 5.97 & 50.39 & 62.42 & 49.51 & 6.06 & 49.63 & 25.98 & N/A & N/A & 45.70 & 100.00 \\ 
\hline
CA & 0.79 & 0.77 & 0.77 & 0.76 & 0.60 & 0.53 & 0.62 & 0.59 & 0.89 & 0.89 & 0.89 & 0.89 \\ 
CD & 0.08 & 0.03 & 0.0 & 0.01 & -0.08 & -0.01 & -0.05 & -0.02 & 0.00 & 0.00 & 0.00 & 0.00 \\ 
    \hline
    \end{tabular}
    \label{tab:xai}
\end{table}

Table~\ref{tab:xai} compares the impact of having users accepting \frank{}'s suggestions randomly (RND) against users deciding on top of \frank{}'s explanations (XAI). 
The first three rows report the percentage of Agreements, Skepticism, and Disagreement between the user and \frank{}.
We notice that they tend to agree, and the disagreement almost always leads to skepticism. 
The fourth and fifth rows show the percentage of the Accepted and Declined \frank{}'s suggestions. 
When XAI is used, we observe a lower agreement rate (Agr) in \texttt{Adult} and \texttt{COMPAS}, but ultimately, looking at the acceptance rate (Acc), these users rely on \frank{} more than their randomizing counterparts, also resulting in a better CD at the expense of CA. 
This confirms that \frank{} is able to provide satisfying explanations to the \textit{Bayesian} and \textit{Similarity} users.
We underline that the \textit{Similarity} expert on \texttt{Adult} is the exception, as they tend to decline.  
Finally, in \texttt{HR}, SLC was never triggered by the \textit{Bayesian}, and only 14 times by the \textit{Similarity} expert (who then declined the 14 suggestions, hence the anomalous percentage).

\section{Conclusion}
\label{sec:conclusion}
We have presented \frank{}, a system based on Skeptical Learning that evolves with the user. 
Compared to traditional SL, \frank{} checks the fairness of the decisions, if they are compliant with external rules, and provides explanations for the suggestions.
Through these additional functions, \frank{} successfully improves the fairness of the datasets and of the model, often outperforming \textsc{SL} in terms of accuracy, especially with less-skilled users. 
Moreover, we noticed that our simulated users accept \frank{}'s explanations most of the time. 
However, at the moment, \frank{} is limited to tabular data and better suitable to those of low dimensionality. 
Future works might extend \frank{} to other data types and decision models, explore alternative stopping conditions, and focus on the \frank{}-user relationships. 
For example, \frank{} could build trust or distrust towards the user, and react accordingly. 
Finally, after being trained in the co-evolutionary process, \frank{}'s model $f$ could be used within a Learning-to-Defer system, with \frank{} making decisions and asking the user when uncertain.

{
\scriptsize
\textbf{Acknowledgment.}
This work is partially supported by the EU NextGenerationEU programme under the funding schemes PNRR-PE-AI FAIR, PNRR-SoBigData.it - Prot. IR0000013, H2020-INFRAIA-2019-1: Res. Infr. G.A. 871042 \textit{SoBigData++}, TANGO G.A. 101120763, and ERC-2018-ADG G.A. 834756 \textit{XAI}. We thank Andrea Bontempelli for his insights on Skeptical Learning.
}

\bibliographystyle{abbrv}
\bibliography{main}

\end{document}